\documentclass[lettersize,journal]{IEEEtran}
\IEEEoverridecommandlockouts
\usepackage{cite}
\usepackage{amsmath,amssymb,amsfonts}
\usepackage{algorithmic}
\usepackage{algorithm}
\usepackage{array}
\usepackage[caption=false,font=normalsize,labelfont=sf,textfont=sf]{subfig}
\usepackage{graphicx}
\usepackage{textcomp}
\usepackage{xcolor}
\usepackage{stfloats}
\usepackage{caption}
\usepackage{booktabs}
\usepackage{multirow}
\usepackage{url}
\usepackage{verbatim}
\usepackage{graphicx}
\usepackage{tcolorbox}
\usepackage{fontawesome}
\usepackage{romannum}
\usepackage[table]{xcolor}
\usepackage{hyperref}
\usepackage{threeparttable}
\usepackage{colortbl}
\usepackage{graphicx}

\def\BibTeX{{\rm B\kern-.05em{\sc i\kern-.025em b}\kern-.08em
    T\kern-.1667em\lower.7ex\hbox{E}\kern-.125emX}}

\begin{document}

\title{\textsc{DISTINCT}: A Description-Guided Branch-Consistency Analysis Framework for Non-Regressive Test Case Generation\\
}

\author{Pengyu Xue\textsuperscript{*},~Yuxiang Zhang\textsuperscript{*},~Zhen Yang\textsuperscript{\dag},~Xiaoxue Ren\textsuperscript{\dag},~Xiang Li,~Pengfei Hu,~Linhao Wu,~Kainan Li
\thanks{\textsuperscript{*}Co-first authors: Pengyu Xue and Yuxiang Zhang.}
\thanks{\textsuperscript{\dag}Corresponding authors: Zhen Yang and Xiaoxue Ren.}
\thanks{Pengyu Xue, Yuxiang Zhang, Zhen Yang, Xiang Li, Linhao Wu, and Kainan Li are with the School of Computer Science and Technology, Shandong University, Qingdao, 266237, China. (e-mail: \{xuepengyu, zhangyuxiang1412\}@mail.sdu.edu.cn,  zhenyang@sdu.edu.cn, \{leexiang, wulinhao, likainan\}@mail.sdu.edu.cn).}
\thanks{Pengfei Hu is with the School of Computer Science and Technology, Shandong University, Qingdao, 266237, China, and the Quancheng Lab, Jinan, 250100, China. (e-mail: phu@sdu.edu.cn)}
\thanks{Xiaoxue Ren is with the Hangzhou High-Tech Zone (Binjiang) Institute of Blockchain and Data Security, Zhejiang University, 310000, Hangzhou, China. (e-mail: xxren@zju.edu.cn).}}

\markboth{IEEE Transactions on Software Engineering,~Vol.~XX, No.~XX, Oct~2025}%
{Zhang \MakeLowercase{\textit{et al.}}: \textsc{DISTINCT}: A Description-Guided Branch-Consistency Analysis Framework for Non-Regressive Test Case Generation}


\maketitle

\begin{abstract}
Automated test-generation research overwhelmingly assumes the correctness of focal methods, yet practitioners routinely face non-regression scenarios where the focal method may be defective.  A baseline evaluation of \textsc{EvoSuite} and two leading Large Language Model (LLM)-based generators, namely \textsc{ChatTester} and \textsc{ChatUniTest}, on defective focal methods reveals that, despite achieving up to 83\% branch coverage, none of the generated tests expose defects, due to a lack of awareness of developer intent. 

To resolve this problem, we first construct two new benchmarks, namely Defects4J-Desc and QuixBugs-Desc, for experiments, where each focal method is equipped with an additional Natural Language Description (NLD) to support code functionality understanding.
Subsequently, we propose \textsc{DISTINCT}, a description-guided branch-consistency analysis framework that transforms LLMs into fault-aware test generators. \textsc{DISTINCT} carries three iterative components: (1) a Generator that derives initial tests based on the NLDs and the focal method, (2) a Validator that iteratively fixes uncompilable tests using compiler diagnostics, and (3) an Analyzer that iteratively aligns test behavior with NLD semantics via branch-level analysis.
Extensive experiments confirm the effectiveness of our approach. Compared to state-of-the-art approaches, \textsc{DISTINCT} achieves an average improvement of 14.64\% in Compilation Success Rate (CSR), 6.66\% in Passing Rate (PR), and particularly 95.22\% in Defect Detection Rate (DDR) across both benchmarks. 
In terms of code coverage, \textsc{DISTINCT} improves Statement Coverage (SC) by an average of 3.77\% and Branch Coverage (BC) by 5.36\%. These results set a new baseline for non-regressive test generation and highlight how description-driven reasoning enables LLMs to move beyond coverage chasing toward effective defect detection.

\end{abstract}


\section{Introduction}

Unit testing \cite{1,2,3} is a cornerstone of software quality assurance. Isolating and exercising discrete units, such as functions or methods, enables early defect detection, prevents issue propagation, and ultimately strengthens software reliability. A well-formed unit test comprises a \emph{setup phase}, which prepares the focal method for execution, and an \emph{assertion phase}, which verifies the observed behavior against the expected outcome \cite{8}.
Writing high-quality tests manually is laborious and error-prone \cite{19,44}. To reduce this burden, researchers have explored a broad spectrum of automated generation techniques, including search-based approaches \cite{4,5,6,7,10,11}, static analyses such as symbolic execution \cite{12,13,14}, dynamic fuzzing \cite{15,16}, and, more recently, Large Language Model (LLM)-based approaches \cite{17,18,19,20,21}.

While these techniques have achieved impressive results, existing evaluations have focused almost exclusively on regression scenarios \cite{22,23,ryan2024code}, in which the focal method is presumed correct and the objective is to detect behavioral deviations caused by subsequent edits. In contrast, developers frequently encounter a more challenging non-regression setting, where the method under test may itself be faulty. Under such conditions, current tools, such as \textsc{EvoSuite}\cite{25} and \textsc{ChatTester}\cite{22}, are easily misled by the incorrect logic of the focal method and generate test cases that reproduce, rather than challenge, the buggy behavior. These methods typically lack access to external semantic guidance (e.g., functional specifications), resulting in test suites that may achieve high structural coverage but fail to reveal actual faults.

\begin{figure}[t]
\centering
\includegraphics[width=0.5\textwidth]{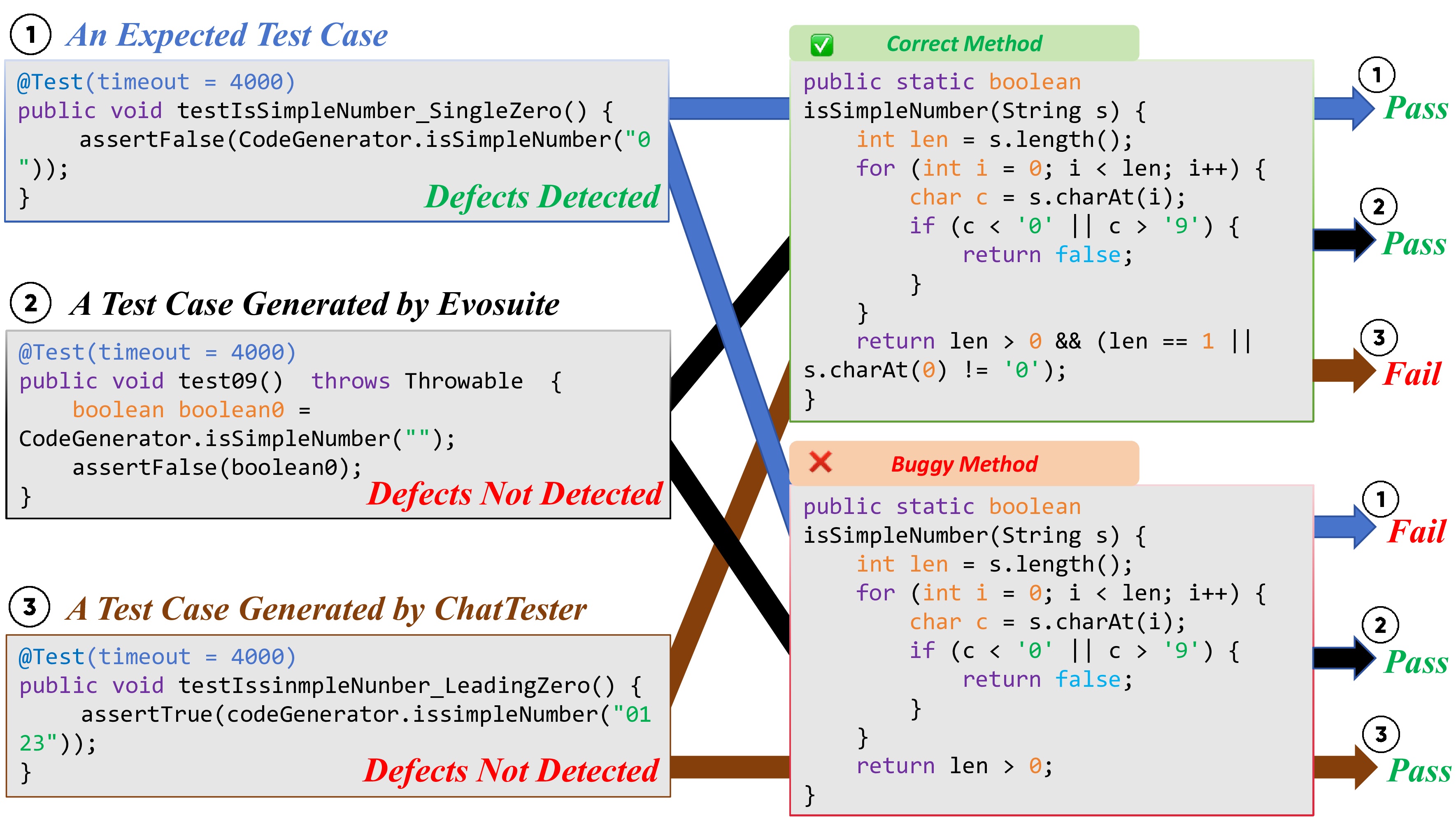}
\caption{Illustrative Example Highlighting Limitations of Existing Testing Techniques}
\label{fig:ills_example}
\end{figure}

Figure \ref{fig:ills_example} demonstrates how the absence of specifications misleads automated tools. In this example, the focal method ``isSimpleNumber'' accepts a string ``s'' and judges whether ``s'' is a simple number.
For the expected test case, the input ``0'' fails on the buggy version of ``isSimpleNumber'' because this version does not handle single-digit zeros, thereby exposing the defect. However, \textsc{EvoSuite} generates a trivial test with an empty string ``" that passes on both versions, failing to reveal the underlying bug. More critically, \textsc{ChatTester} generates ``0123'' as a test input, which passes on the buggy version of ``isSimpleNumber'' but fails on its correct version. Because these tools treat the buggy implementation as the ground truth, they inadvertently reinforce incorrect behaviors like passing with ``0123'' rather than rejecting them as a bug. As detailed in Section \ref{Empirical study}, without semantic cues, such tools are 100\% misled by buggy focal methods.

 
To systematically illustrate invalid test cases affected by the misleading behaviors of focal methods like those above, we first construct a benchmark comprising 441 defective Java methods sourced from Defects4J \cite{26} (version v2.0.0), and 40 buggy cases from QuixBugs \cite{51}, which is detailed in Subsection \ref{Dataset Construction}. 
Subsequently, we compare the ability of three representative testing techniques, including a search-based approach (i.e., \textsc{EvoSuite} \cite{25}), and two recent LLM-based approaches (i.e., \textsc{ChatTester} \cite{22} and \textsc{ChatUniTest} \cite{23}), in identifying bugs on the constructed benchmarks. 
Experimental results show that \textsc{EvoSuite} attains high scores of 83.29\%-95.67\% and 73.57\%-96.89\% on Branch Coverage (BC) and Statement Coverage (SC) across Defects4J and QuixBugs, yet still fails to expose any faults in the focal methods.
\textsc{ChatTester} and \textsc{ChatUniTest} perform similarly; they can generate human-readable and high-coverage test suites, but their Defect Detection Rate (DDR) remains stuck at 0.00\%. This highlights a key insight of our work: \textbf{being misled by focal methods, even achieving high coverage, auto-generated test cases cannot guarantee meaningful bug identification}.


\vspace{2mm}
\vspace{-1mm}

In fact, in real-world development, programmers typically need to write comments for code snippets in Natural-Language Descriptions (NLDs) format, thereby conveying functional specifications. Therefore, providing such NLDs for test case generation algorithms is a natural and feasible behavior to address the above limitations.
Hence, based on the previously extracted samples from Defects4J and QuixBugs, we extend them with human-crafted NLDs, thereby forming two new benchmarks, namely \textbf{Defects4J-Desc} and \textbf{QuixBugs-Desc}, which are constructed following the procedure detailed in Subsection~\ref{Benchmark Construction}, to fit our experimental setting of non-regression testing.
Subsequently, we propose \textbf{\textsc{DISTINCT}}, a \textbf{\underline{D}}escription-gu\textbf{\underline{I}}ded branch-con\textbf{\underline{S}}is\textbf{\underline{T}}ency analys\textbf{\underline{I}}s framework for \textbf{\underline{N}}on-regressive test \textbf{\underline{C}}ase genera\textbf{\underline{T}}ion. \textsc{DISTINCT} marries the generative power of LLMs with a branch-level semantic-refinement pipeline consisting of three stages: (1) \emph{Generator}: generating initial tests from the NLD and focal method, (2) \emph{Validator:} repairing compilation errors through iterative compiler feedback, and (3) \emph{Analyzer:} aligning test logic with the described intent via branch-consistency analysis that revises or extends incorrect assertions.

We implement \textsc{DISTINCT} and evaluate it on Defects4J-Desc and QuixBugs-Desc. Compared to state-of-the-art approaches, our framework improves Compilation Success Rate (CSR) by 19.6\% and 9.68\%, and Passing Rate (PR) by 9.32\% and 4.17\% across two benchmarks, respectively. More importantly, it achieves a 149.26\% and 41.18\% gain in DDR, while enhancing BC by 6.34\% and 4.37\%, and SC by 2.07\% and 5.48\%.
Ablation studies reveal that both the \emph{Validator} and \emph{Analyzer} components, as well as their internal designs, independently enhance test case generation performance in the non-regression setting.
Besides, discussion experiments confirm \textsc{DISTINCT}’s generality across diverse LLM sizes and architectures, and additional hyper-parameter analysis showcases the performance variation of \textsc{DISTINCT} under various settings, providing straightforward suggestions for developers during their practical usage.
This paper makes the following contributions:

\begin{itemize}
\item We lead the first in-depth study of unit-test generation in the non-regression setting, where the focal method may be faulty.
\item We construct two enriched benchmarks \cite{55, 51}, namely \textbf{Defects4J-Desc} and \textbf{QuixBugs-Desc}, annotated with NLDs and focal methods with paired buggy/fixed versions.
\item We propose \textsc{DISTINCT}, a novel LLM-based unit test generation framework that integrates focal methods and NLDs through branch-consistency analysis.
\item Extensive experiments demonstrate the effectiveness of \textsc{DISTINCT} against state-of-the-art approaches, including ablation studies, discussion experiments, and hyper-parameter analysis. 
\end{itemize}

\section{Empirical study}
\label{Empirical study}
This empirical study aims to resolve our first Research Question (RQ1): \textbf{How effectively do existing automated test
generation techniques identify defects in non-regression testing scenarios? }
To answer this question, we conduct experiments following the setup below.

\subsection{Approaches}
To address RQ1, we consider four typical approaches: \textsc{EvoSuite} \cite{25}, individual LLM, and two LLM-augmented approaches, namely \textsc{ChatTester} \cite{22}, \textsc{ChatUniTest} \cite{23}.

(1) \textsc{EvoSuite}\cite{25} is a widely used automated test generation tool for Java, based on search-based techniques\cite{4, 10, 11}. It aims to maximize code coverage (e.g., BC or SC) by generating JUnit test suites through evolutionary algorithms. \textsc{EvoSuite} serves as a strong traditional approach and is considered one of the SOTA representatives in traditional test generation, often used in software testing \cite{tang2024chatgpt,22} and program repair \cite{ruan2024evolutionary,motwani2020quality} studies.

(2) We also involve an individual LLM for experiments because we intend to understand the performance improvement of current LLM-augmented approaches in a brand-new testing configuration, i.e., non-regression testing.  

(3) \textsc{ChatTester} \cite{22} is an LLM-augmented test generation framework that decomposes the process into two stages: it first infers the intended functionality of the focal method through an intention prompt, and then generates a corresponding unit test using a generation prompt that incorporates both the inferred intention and code context. To improve test compilability, it applies an iterative refinement process that analyzes compilation errors and relevant code context, updating the prompt until a valid test is produced or a maximum iteration limit is reached.

(4) \textsc{ChatUniTest} \cite{23} is another LLM-augmented framework for test generation, which first parses the code to extract structural and dependency information, then constructs adaptive prompts using context-aware templates. Then, the LLM generates complete test classes, which are validated through syntax, compilation, and runtime checks. If errors occur, \textsc{ChatUniTest} applies rule-based and LLM-based repair strategies to iteratively refine the tests.

\subsection{Dataset Construction}
\label{Dataset Construction}
Most existing benchmark datasets for unit testing are primarily applicable to regression testing scenarios, where the focal method is assumed to be logically correct. As such, they are not suitable for non-regression testing. To experiment, we select two widely used program repair benchmarks, namely Defects4J (version v2.0.0) and QuixBugs, for non-regression testing dataset construction.
Nonetheless, most defects in Defects4J span multiple methods, classes, or even packages, and thus do not typically correspond to a single, self-contained function suitable for unit test generation. Moreover, many methods rely on complex contexts, configuration files, databases, or external states, which makes them less ideal as purely side-effect-free units. To address this, we filter out complex defects that require interprocedural context or cross-method interactions. Finally, we retained 441 buggy methods from Defects4J and refer to this subset as Defects4J-Sub. Defects in Defects4J-Sub can only be triggered or observed through a single focal method. In contrast, the 40 Java programs in QuixBugs are relatively independent and logically simple, making QuixBugs an appropriate dataset for our experiments.

\subsection{Implementation}
\label{Implementation}
Each tool is configured with appropriate settings to ensure a fair and reproducible comparison on the two non-regressive testing datasets constructed above.



For \textsc{EvoSuite}, we set a random seed, the search budget to 300 seconds (\texttt{-Dsearch\_budget=300}), and the memory limit to 1500 MB (\texttt{-mem 1500}). Additionally, we configured the tool with \texttt{assertion\_strategy = MUTATION} and \texttt{assertion\_timeout = 60s} to enhance assertion generation. Using this configuration, \textsc{EvoSuite} is employed to generate corresponding unit tests for the focal methods.

For the individual LLM, we use the Deepseek-v3 \cite {21} model. The design of the basic prompt follows the methodology adopted in the study of \textsc{ChatTester}\cite{22}, thereby ensuring consistency with established practices. To be specific, the prompt consists of two components: (1) a Natural Language (NL) part, which includes a role-playing instruction and a task description to guide LLMs in generating unit tests; and (2) a Code Context (CC) part, which provides the focal method along with relevant class-level context, including fields and method signatures. A specific use case of an Individual LLM is presented in \cite{55}.

As for the other two LLM-augmented approaches, we adopt DeepSeek as the backbone model for experiments as well, thereby ensuring a fair comparison. Apart from this, other settings are consistent with those in their original papers \cite{22, 46}. In addition, we extract the first five generated test cases for each LLM in the subsequent experiments.
To ensure a fair comparison, we likewise select only the first five test cases produced by \textsc{EvoSuite} for evaluation, even though \textsc{EvoSuite} is inherently designed to generate an unbounded number of tests through evolutionary search. The impact of the number of generated test cases on the performance of LLM-based test generation tools is further examined in Subsection \ref{RQ5}.


\subsection{Evaluation Metrics} 
We assess test generation quality using the evaluation metrics below:

\begin{itemize}
    \item \textbf{Compilation Success Rate (CSR)}: The percentage of generated test cases that can be successfully compiled without syntax errors.
    
    \item \textbf{Passing Rate (PR)}: The percentage of compiled test cases that can run to completion without runtime failures on focal methods of fixed versions.
    
    \item \textbf{Defect Detection Rate (DDR)}: The capability to reveal known defects in the benchmark, measured by the ratio of detected defects to total known defects.
    
    \item \textbf{Branch Coverage (BC)}: The percentage of conditional branches in the fixed method that are executed by the generated tests.
    
    \item \textbf{Statement Coverage (SC)}: The percentage of executable statements in the fixed method that are exercised by the tests.
\end{itemize}

\subsection{Results Analysis}





\begin{table*}[htbp]
\centering
\footnotesize
\renewcommand{\arraystretch}{1.2}
\setlength{\abovecaptionskip}{1pt}
\setlength{\tabcolsep}{10pt} 
\caption{Comparison of test generation tools on Defects4J-Sub and QuixBugs across diverse evaluation metrics.}
\begin{tabular}{lcccccccccc}
\toprule
\multirow{2}{*}[-0.8em]{\textbf{Baselines}} & \multicolumn{5}{c}{\textbf{Defects4J-Sub}} & \multicolumn{5}{c}{\textbf{QuixBugs}} \\
\cmidrule(lr){2-6} \cmidrule(lr){7-11}
& CSR& PR & DDR & BC & SC
& CSR & PR & DDR & BC & SC \\
\cmidrule(lr){1-6} \cmidrule(lr){7-11}
\textsc{EvoSuite} & \textbf{97.10\%} & \textbf{78.40\%} & 0\% & 32.46\% & 36.68\% & \textbf{100.0\%} & \textbf{100.0\%} & 0\% & \textbf{95.25\%} & \textbf{96.68\%} \\
Individual LLM & 22.40\% & 5.44\% & 0\% & 58.40\% & 68.90\% & 75.00\% & 37.50\% & 0\% & 86.73\% & 76.59\% \\
\textsc{ChatTester} & 22.95\% & 3.28\% & 0\% & \textbf{64.60\%} & \textbf{79.75\%} & 77.50\% & 52.5\% & 0\% & 87.65\% & 77.75\% \\
\textsc{ChatUniTest} & 20.19\% & 7.4\% & 0\% & 58.30\% & 67.80\% & 75.00\% & 50.00\% & 0\% & 87.35\% & 77.40\% \\
\bottomrule
\end{tabular}
\label{table1}
\end{table*}

Table~\ref{table1} presents a comprehensive comparison of four unit test generation tools in terms of each evaluation metric.
As for CSR and PR, \textsc{EvoSuite} demonstrates consistently strong performance. On Defects4J-Sub and QuixBugs, their test cases compile successfully at rates of 97.10\% and 100.00\%, and achieve test passing rates of 78.40\% and 100.00\%, respectively. The reason is that \textsc{EvoSuite} executes the generated tests against the focal method and only retains those that run without exceptions, effectively filtering out invalid or infeasible inputs. Since we evaluate the generated tests on the fixed versions of focal methods, which are unseen to \textsc{EvoSuite}, their success rates on CSR and PR do not always reach 100\%.
\textsc{ChatTester}, which utilizes LLMs with iterative refinement, performs weaker, achieving 22.95\% in terms of CSR and 3.28\% in terms of PR on Defects4J-Sub, but improving to 77.50\% and 52.50\% on QuixBugs. \textsc{ChatUniTest}, following a similar generate-then-repair paradigm, performs neck-to-neck with \textsc{ChatTester}, being relatively higher than it on Defect4J-sub while lower on the Quixbugs. Although they both outperform the individual LLM overall, but perform weaker than \textsc{EvoSuite}, owing to the lack of a filtering mechanism equipped in \textsc{EvoSuite}. The performance discrepancies between the two benchmarks are due to their different complexity, where Defect4J-Sub is composed of real-world programs with higher complexity, while QuixBugs only contains single-line bugs, which are relatively easier.

For coverage analysis, Table~\ref{table1} reports statement and branch coverage based only on executable test cases (i.e., those that compile and run). \textsc{EvoSuite}, although optimized for coverage using genetic algorithms, attains only 32.46\% on SC and 36.88\% on BC towards Defects4J-Sub, significantly lagging behind two LLM-based approaches. 
As for QuixBugs, it obtains the highest performance among its counterparts, reaching 95.25\% and 96.68\% on BC and SC, respectively. One major reason is that \textsc{EvoSuite} typically generates only a single assertion per test, and the dependent generic algorithm evolves relatively slowly to generate effective test cases with limited attempts for complex focal methods, such as samples in Defects4J-Sub, compared with LLM-based approaches with powerful analysis capabilities. 

Despite promising results in coverage and execution correctness, a key limitation emerges in non-regression testing: all tools achieve 0\% of DDR on both Defects4J-Sub and QuixBugs, as shown in Table~\ref{table1}. The reason is apparent that all tools under experiment lack access to the intended behavior of focal methods, but distort the original intent by the buggy logic of focal methods, making it impossible to detect the true bugs.


\begin{tcolorbox}[colback=gray!10!white, colframe=gray!50!black, boxrule=0.5pt, before skip=10pt, after skip=10pt]
\textbf{\faPencil~Finding 1}: Although existing unit test generation approaches achieve commendable performance in terms of execution correctness or test coverage, they fail to generate defect-revealing tests, owing to the lack of intents of focal methods.


\end{tcolorbox}

\section{\textsc{DISTINCT}}

\begin{figure}[h]
    \centering
\includegraphics[width=0.5\textwidth]{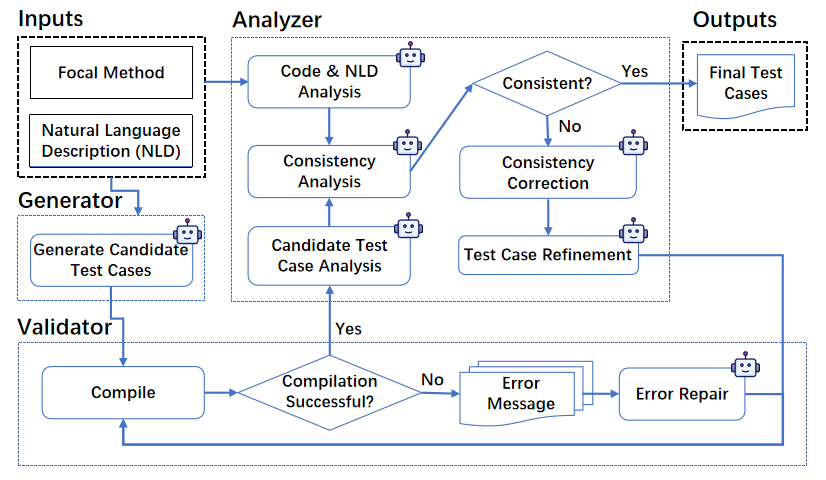} 
    \caption{The Workflow of \textsc{DISTINCT}}
    \label{fig:workflow}
\end{figure}

Based on the findings above, we propose a novel LLM-based unit test generation framework for non-regression testing, namely \textsc{DISTINCT}. 
\textsc{DISTINCT} introduces the Natural Language Description (NLD) of code functionalities as crucial extra information to significantly enhance the generation of defect-revealing test cases. \textsc{DISTINCT} consists of three main stages, including (1) Generator: generating candidate test cases from the focal method and its NLD, (2) Validator: validating and repairing these test cases through iterative compilation, and (3) Analyzer: iteratively refining test semantics by aligning their logic with the behavior described in the NLD. The workflow of \textsc{DISTINCT} is illustrated in Figure~\ref{fig:workflow}.

\begin{figure}[h]
    \centering
    \includegraphics[width=0.5\textwidth]{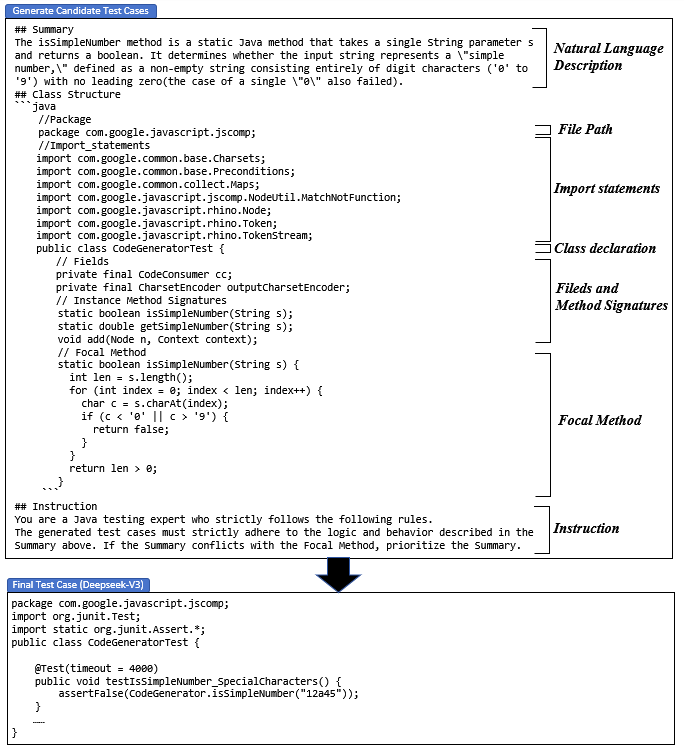} 
    \caption{Generate Candidate Test Cases of \textit{isSimpleNumber}}
    \label{fig:generateprompt}
\end{figure}

\subsection{Generator}

Given the additional NLD in our experimental setting, we extended the basic prompt template originally used in \textsc{ChatTester} and accordingly modified a series of instructions.
The resulting prompt includes the focal method and its associated NLD depicting functionalities, along with other contextual 
information such as a class declaration, related method signatures, fields, and import statements. Generator also includes defective focal methods as one of the inputs because they consist of low-level implementation details, and LLMs tend to overlook certain execution paths if they rely solely on the highly abstractive NLDs. 
Given the above prompt, an LLM is then guided to generate \textit{candidate test cases} for the defective \textit{focal method}. Figure~\ref{fig:generateprompt} illustrates a generation example for the focal method \textit{isSimpleNumber}.

\subsection{Validator}
Test case repair is a commonly used module owing to its verified effectiveness \cite{59, 22, 23}. Therefore, we inherit the Validator of \textsc{ChatTester} \cite{22} with compile and error repair phases in \textsc{DISTINCT}. 
This repair process is performed iteratively. If \textit{candidate test cases} continue to fail compilation after a repair attempt, the Validator re-extracts the latest error messages and triggers another repair iteration. This loop continues until the test case compiles successfully or the number of iterations reaches a determined limit; in this work, we initially set this value to 5.

\subsection{Analyzer}
For \textit{candidate test cases} that pass compilation, we employ an Analyzer to optimize their test coverage and defect-revealing ability. 
Analyzer contains five modules, namely \textit{Code \& NLD Analysis}, \textit{Candidate Test Case Analysis}, \textit{Consistency Analysis}, \textit{Consistency Correction}, and \textit{Test Case Refinement}
for progressively revealing potentially uncovered branches and unreached defects in \textit{candidate test cases}.

\subsubsection{Code \& NLD Analysis}
To extract accurate logical branches from the \textit{focal method}, we perform two rounds of logic analysis using the LLM. In the first round (Code Analysis), we prompt the LLM to analyze the control flow structure of the \textit{focal method} to identify all possible execution paths. Specifically, we construct a prompt template defined as: ``\#\# Focal Method\verb|\n|$\{$FOCAL\_METHOD$\}$\verb|\n|\#\# Instruction\verb|\n|You are a Senior Java Control Flow Analyst. Analyze the following focal method to identify all possible logical branches based on its control flow structure. Focus on static code analysis: extract every execution path, including all conditional branches, loops, and edge cases. Follow IEEE Structured Testing Guidelines with attention to full path coverage.'', where $\{$FOCAL\_METHOD$\}$ is the placeholder for a focal method. This prompt guides the LLM in performing a static, structure-based analysis to extract the initial set of logical branches.

In the second round (NLD Analysis), we enhance the analysis by incorporating the semantic understanding provided by the NLD. Based on this NLD, we prompt the LLM to revise and refine the previously extracted branches. The prompt template in this stage is defined as: ``$\{$CODE\_ANALYSIS\_OUTPUTS$\}$\verb|\n|\#\# Summary\verb|\n|$\{$SUMMARY$\}$\verb|\n|\#\# Instruction\verb|\n|You are a Java Logic Correction Expert. Given the Summary of the focal method and its previously extracted logical branches, identify and correct any incorrect or incomplete branches. Ensure the final logical branches align precisely with the described intention. Add any missing branches and remove or fix semantically incorrect ones.'', where $\{$CODE\_ANALYSIS\_OUTPUTS$\}$ and $\{$SUMMARY$\}$ are the placeholders of the outputs of code analysis and the NLD.  This analysis enables the correction of erroneous logical branches and the identification of any branches missed during the initial code analysis.

\begin{figure}[h]
    \centering
    \includegraphics[width=0.5\textwidth]{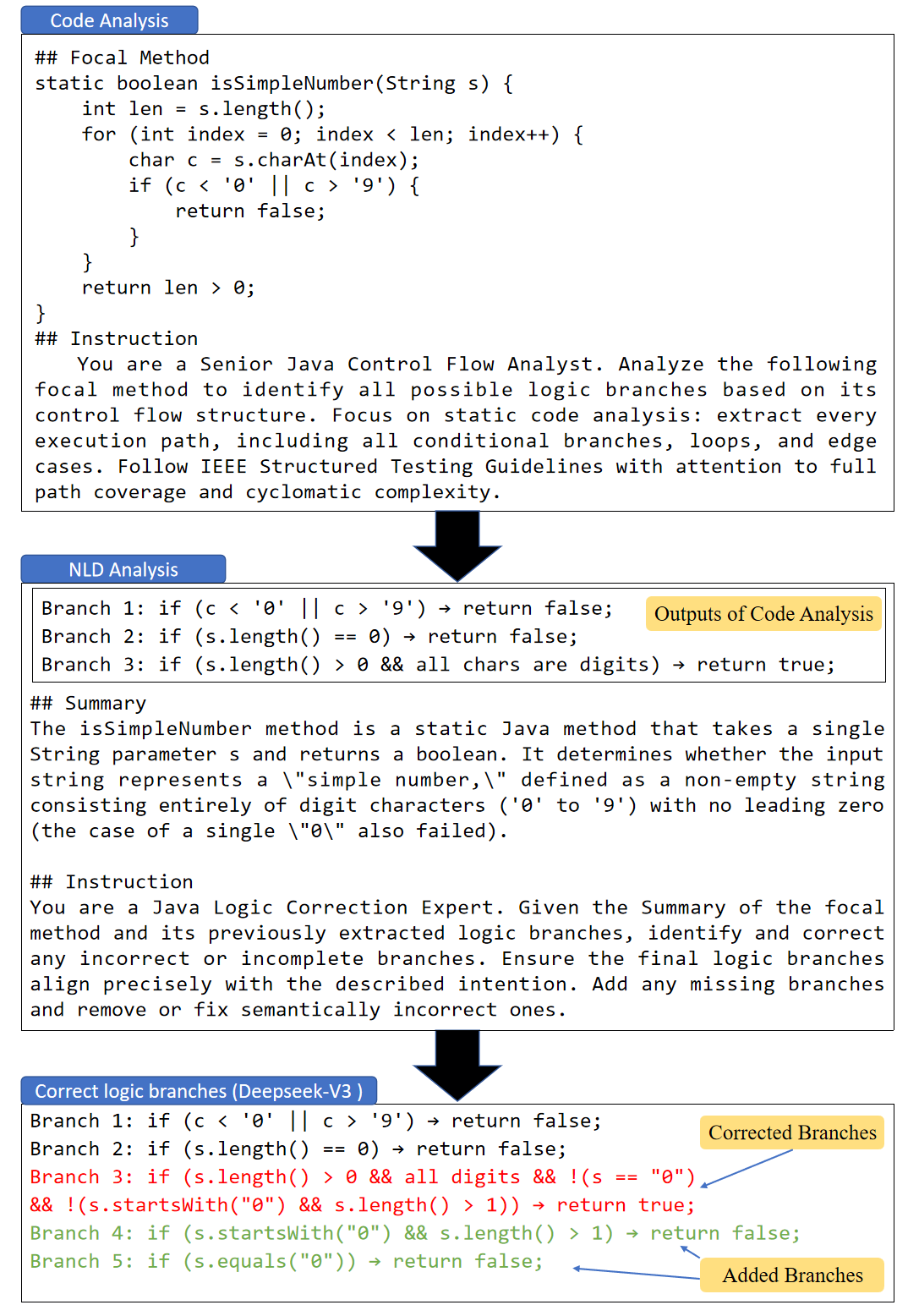} 
    \caption{Code \& Description Analysis of \textit{isSimpleNumber}}
    \label{fig:example1}
\end{figure}

Figure~\ref{fig:example1} illustrates a typical example of the focal method \textit{isSimpleNumber}. Initially, LLM identifies three control flow branches based on the focal method. However, Branch 3 fails to account for edge cases, such as strings starting with a leading zero (e.g., ``01''), which deviates from the original intent of the method \textit{isSimpleNumber}. By incorporating the NLD of method \textit{isSimpleNumber}, Branch 3 is rectified. Besides, other missing branches are also added, thereby aligning precisely with the intended behavior. We refer to the final refined logic as the \textit{correct logical branches}, serving as a pseudo-reference for the follow-up steps.

\subsubsection{Candidate Test Case Analysis}
\begin{figure}[h]
    \centering
    \includegraphics[width=0.4\textwidth]{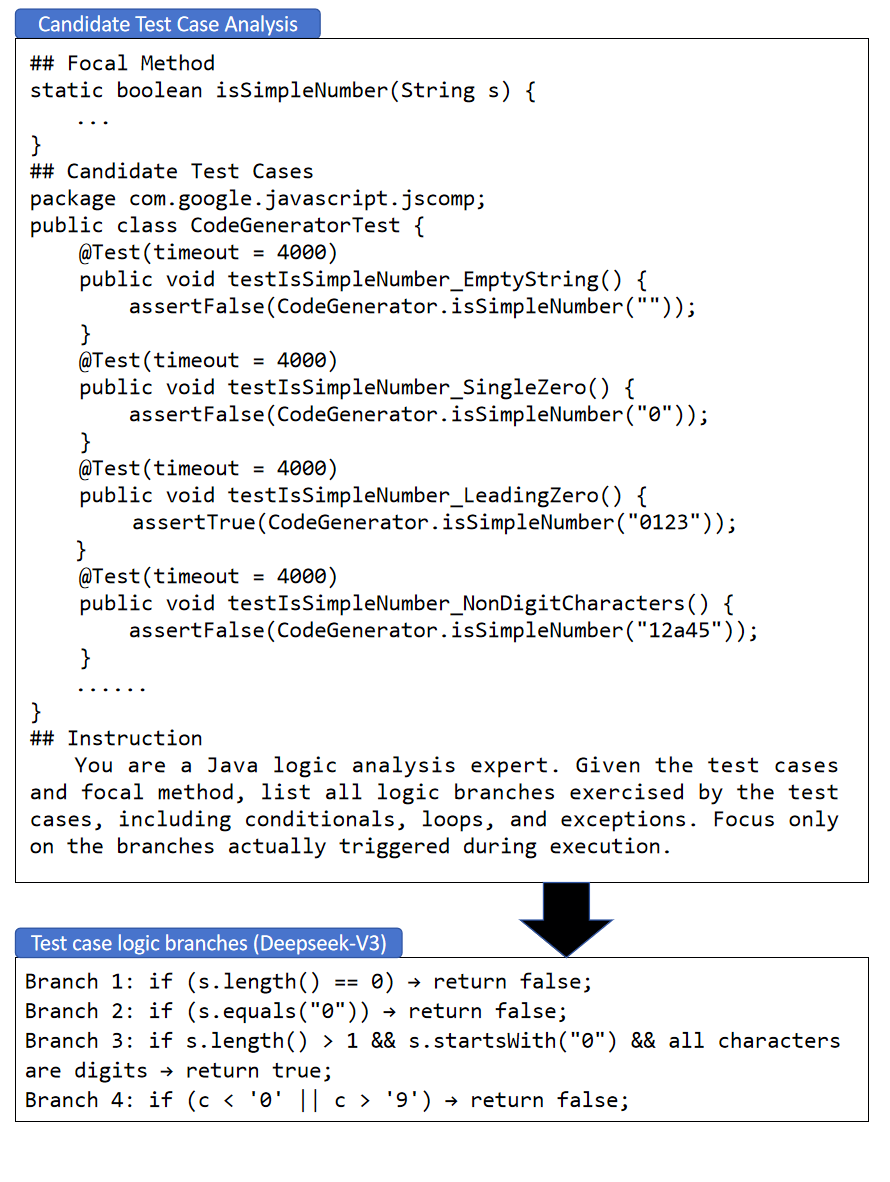} 
    \caption{Candidate Test Case Analysis of \textit{isSimpleNumber}}
    \label{fig:example2}
\end{figure}

In this stage, we perform logic analysis on the compilation-passed \textit{candidate test cases} to extract all their logical branches, referred to as the \textit{test case logical branches}. To guide this analysis, we include \textit{candidate test cases} and the associated focal method to construct the following prompt template: ``\#\# Focal Method\verb|\n|$\{$FOCAL\_METHOD$\}$\verb|\n|\#\# Candidate Test Cases\verb|\n|$\{$CANDIDATE\_TESTS$\}$\verb|\n|\#\# Instruction\verb|\n|You are a Java logic analysis expert. Given the test cases, list all logical branches exercised by the test cases, including conditionals, loops, and exceptions. Focus only on the branches actually triggered during execution.'', where $\{$CANDIDATE\_TESTS$\}$ is the placeholder of \textit{candidate test cases}. This prompt directs the LLM to identify only the branches that are actively covered by the execution of the \textit{candidate test cases}. An example is shown in Figure~\ref{fig:example2}.

\subsubsection{Consistency Analysis}
\begin{figure}[h]
    \centering
    \includegraphics[width=0.4\textwidth]{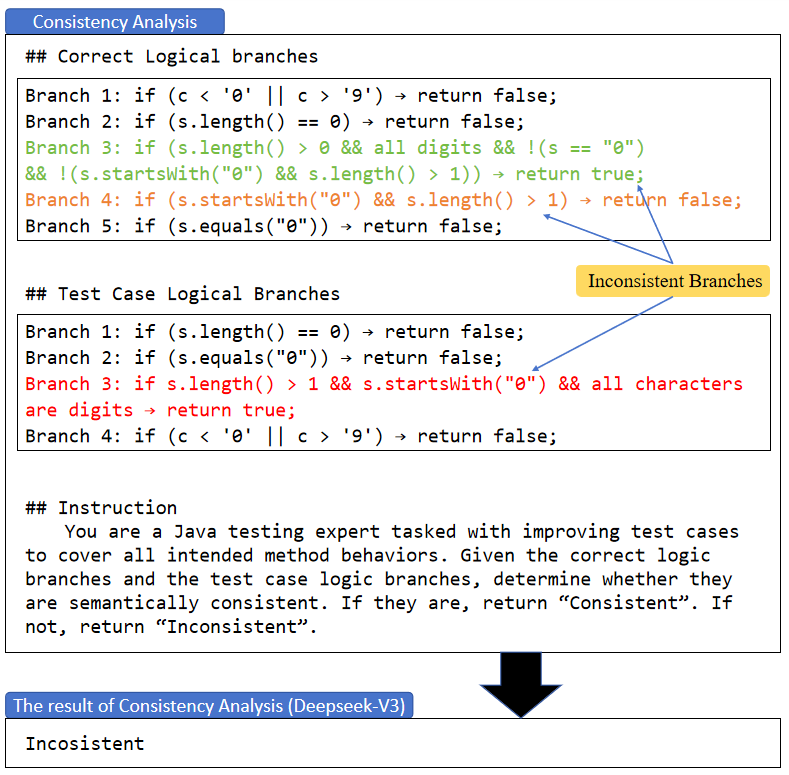} 
    \caption{Consistency Analysis of \textit{isSimpleNumber}}
    \label{fig:analysisagree}
\end{figure}

Finally, we construct the following prompt template: ``\#\# Correct Logical Branches\verb|\n|$\{$CORRECT\_BRANCH$\}$\verb|\n|\#\# Test Case Logical Branches\verb|\n|$\{$TEST\_CASE\_BRANCH$\}$\verb|\n|\#\# Instruction\verb|\n|You are a Java testing expert tasked with improving test cases to cover all intended method behaviors. Given the correct logical branches and the test case logical branches, determine whether they are semantically consistent. If they are, return \textbackslash``Consistent\textbackslash''. If not, return \textbackslash``Inconsistent\textbackslash''.'', where $\{$CORRECT\_BRANCH$\}$ and $\{$TEST\_CASE\_BRANCH$\}$ are placeholders of \textit{correct logical branches} and \textit{test case logical branches}.
We use this prompt to compare the above two branch sets. If the result is ``Consistent'', the \textit{candidate test cases} is considered to meet our criteria and is likely to be an effective \textit{final test} capable of detecting the defect. If the result is ``Inconsistent'', a \textit{consistency correction} is required to further refine the test suite. A specific example is shown in Figure~\ref{fig:analysisagree}.

\subsubsection{Consistency Correction}

\begin{figure}[h]
    \centering
    \includegraphics[width=0.4\textwidth]{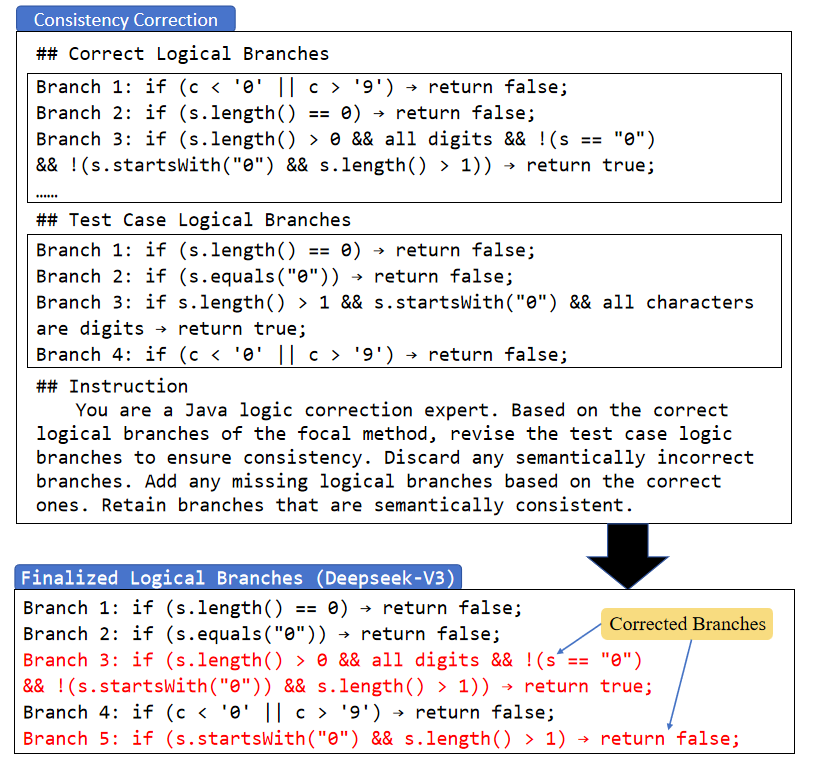} 
    \caption{Consistency Correction of \textit{isSimpleNumber}}
    \label{fig:branchconsistent}
\end{figure}

To maintain the mapping relations between \textit{candidate test cases} and \textit{test case logical branches} derived from it, this stage corrects faulty branches in \textit{test case logical branches} according to the \textit{correct logical branches}. 
Specifically, for branches in the \textit{test case logical branches} that deviate semantically from the \textit{correct logical branches}, we hope the LLM discards them and replaces with the correct ones. Missing branches are similarly supplemented using the \textit{correct logical branches}. Logical branches that are semantically consistent across both are retained, thereby obtaining the \textit{finalized logical branches}.
To guide this correction process, we prompt the LLM with the following template: ``\#\# Correct Logical Branches\verb|\n|$\{$CORRECT\_BRANCH$\}$\verb|\n|\#\# Test Case Logical Branches\verb|\n|$\{$TEST\_CASE\_BRANCH$\}$\verb|\n|\#\# Instruction\verb|\n|You are a Java logic correction expert. Based on the correct logical branches of the focal method, revise the test case logic branches to ensure consistency. Discard any semantically incorrect branches. Add any missing logical branches based on the correct ones. Retain branches that are semantically consistent.''. An illustrative example is shown in Figure~\ref{fig:branchconsistent}.

\subsection{Test Case Refinement}
\begin{figure}[h]
    \centering
    \includegraphics[width=0.4\textwidth]{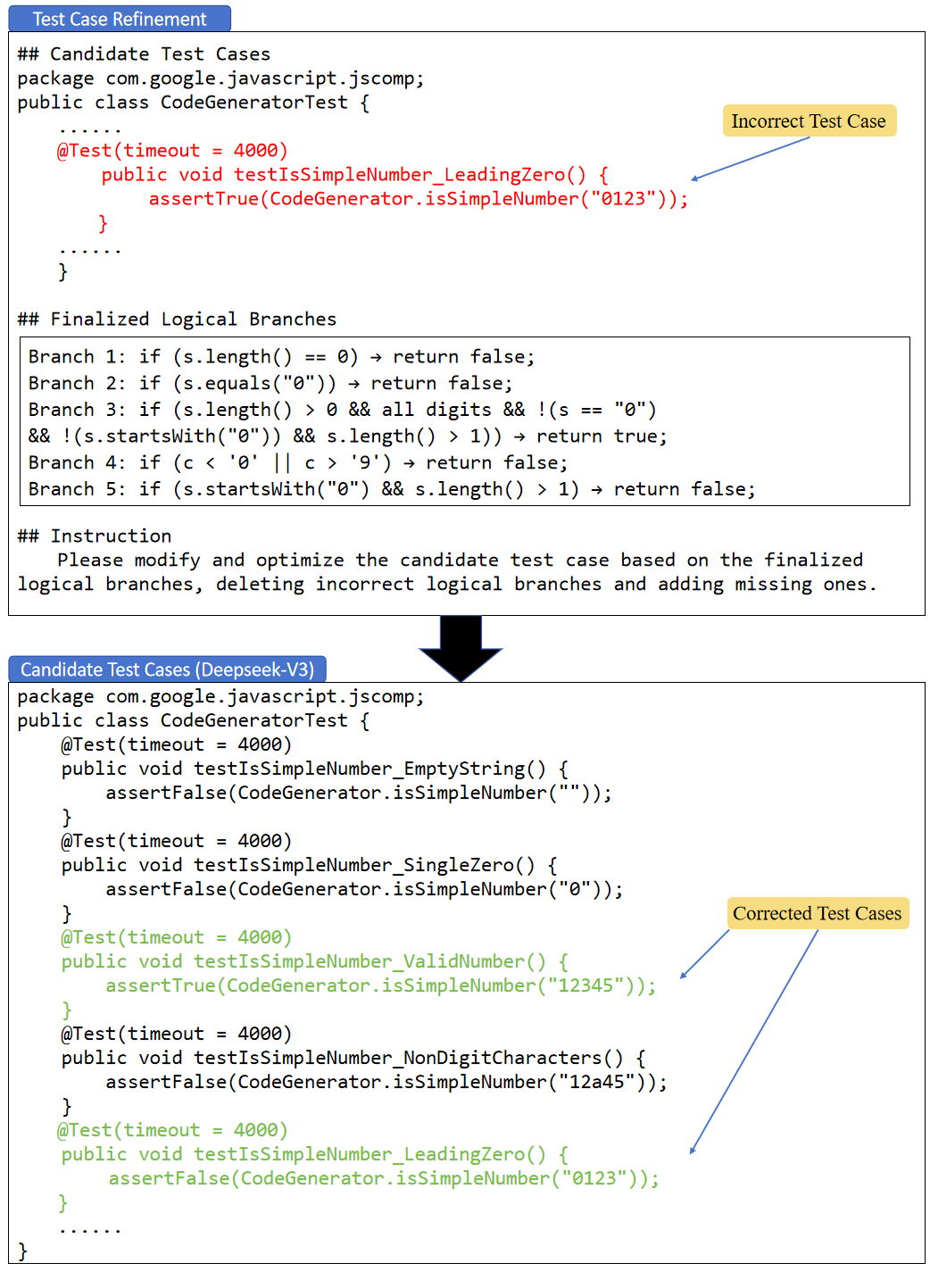} 
    \caption{Test Case Refinement of \textit{isSimpleNumber}}
    \label{fig:iterativegeneration}
\end{figure}

Guided by the \textit{finalized logical branches}, we iteratively prompt the LLM to generate additional \textit{candidate test cases}, and the LLM’s output is then fed into the Validator to initiate a new iteration. A specific example is shown in Figure~\ref{fig:iterativegeneration}. This loop continues until the \textit{candidate test cases} reach the consistency result, or the number of iterations reaches a predefined limit; in this work, we empirically set this value to 5.

\section{Evaluation}
\subsection{Benchmark Construction}\label{Benchmark Construction}

While both Defects4J-Sub and QuixBugs are applicable in non-regression testing contexts, they lack Natural Language Descriptions (NLDs) of focal methods to aid in understanding their intended behavior.
To better evaluate our proposed test case generation framework, \textsc{DISTINCT}, we extend Defects4J-Sub and QuixBugs with manually crafted NLDs, constructing Defects4J-Desc and QuixBugs-Desc.
To simulate how developers comprehend focal methods, we engaged five professional Java engineers, each with over three years of development experience, to independently write textual NLDs for the fixed versions of focal methods. This process ensures that the NLDs accurately reflect the original intent of the code.

To ensure the objectivity, consistency, and informativeness of each NLD, we establish a standardized annotation protocol consisting of 3 principles \cite{mcconnell2004code,ljung2022clean,meyer1997object}:

\textbf{(1) Functional Abstraction:} The first sentence concisely describes the method’s primary purpose and its behavioral role in the enclosing class. It follows the pattern ``The \texttt{[method name]} method in the \texttt{[class name]} class \texttt{[does something]}'', where \texttt{[method name]}, \texttt{[class name]}, and \texttt{[does something]} are placeholders.

\textbf{(2) Input/Output Specification:} The NLD clearly explains the types and semantic roles of all input parameters and return values, including any noteworthy assumptions such as non-null requirements or exception conditions.

\textbf{(3) Core Logic Extraction:} The central logic of the method is paraphrased in natural language, capturing key control flow, calculations, and use of standard libraries (e.g., \texttt{Math.sqrt}, \texttt{Erf.erf}). Rather than line-by-line translation, the NLD conveys what the method computes and how.


Each NLD is independently annotated by two engineers and finalized through reconciliation and proofreading. As a result, our Defects4J-Desc comprises 441 focal methods from 17 Java projects, and QuixBugs-Desc contains 40 Java programs, with each sample containing its buggy version, fixed version, and a high-quality textual NLD describing its fixed version’s behavior. To the best of our knowledge, this is the first benchmark explicitly designed for unit testing of defective focal methods. It bridges a critical gap in test generation research, providing a realistic and rigorous foundation for evaluating test case generation tools under non-regressive scenarios. A detailed statistical overview of the two datasets, Defects4J-Desc and QuixBugs-Desc, including the number of bugs, as well as length statistics of both code and NLDs in tokens, is presented in Table \ref{tab:stats}.


\begin{table}[htbp]
    \centering
    \setlength{\abovecaptionskip}{1pt}
    \caption{Statistics of the Evaluation Datasets}
    \label{tab:stats}
    \begin{tabular}{c c c c c} 
    \toprule
    \multirow{2}{*}{\textbf{Dataset}} & 
    \multirow{2}{*}{\textbf{Metric}} & 
    \textbf{Buggy} & 
    \textbf{Fixed} & 
    \textbf{NLD} \\
    & & \textbf{(Tokens)} & \textbf{(Tokens)} & \textbf{(Tokens)} \\
    \midrule
    \multirow{5}{*}{\parbox{2.5cm}{\centering \textbf{Defects4J-Desc} \\ (\# Bugs: 441) \\}}
    & $\text{Max.}$ & 1491 & 1479 & 299 \\
    & $\text{Min.}$ & 50 & 51 & 50 \\
    & $\text{Med.}$ & 194 & 251 & 85 \\
    & $\text{Avg.}$ & 282.5 & 327.2 & 94.9  \\ 
    & $\text{S.D.}$ & 248.3 & 272.6 & 39.3 \\ 
    \midrule
    \multirow{5}{*}{\parbox{2.5cm}{\centering \textbf{QuixBugs-Desc} \\ (\# Bugs: 40)}}
    & $\text{Max.}$ & 424 & 457 & 70 \\
    & $\text{Min.}$ & 47 & 47 & 47 \\
    & $\text{Med.}$ & 143 & 149 & 57 \\
    & $\text{Avg.}$ & 163.6 & 167.1 & 56.7\\ 
    & $\text{S.D.}$ & 77.9 & 79.9 & 5.4 \\ 
    \bottomrule
    \end{tabular}
\end{table}

\subsection{Evaluation Setup}\label{Evaluation Setup}
Below introduces the rest of the RQs we explored. 
LLMs, participating in experiments below, follow the same settings as mentioned in Section \ref{Implementation} for implementation. All experiments are carried out on the benchmarks constructed above in Section \ref{Benchmark Construction}, namely Defects4J-Desc and QuixBugs-Desc.

To minimize the inherent stochasticity of LLMs during inference, we fixed the \textit{temperature} to 0. This configuration ensures that the LLMs’ randomness is reduced to a minimum, thereby ensuring the reproducibility of the experiments.
Nevertheless, a minor degree of randomness may still remain. We further evaluated the stability of our results by repeating a subset of experiments three times using the basic prompt configuration under the same experimental conditions. The subset comprised a number of focal methods from Defects4J-Desc and QuixBugs-Desc.
Across repetitions, the observed variance of all LLMs under test in key metrics (CSR, PR, and DDR) was consistently below 5\%, indicating that randomness exerts negligible influence on the reported conclusions.

\textbf{RQ2 (Overall Effectiveness): How does DISTINT perform against other approaches?} 
We modify baseline approaches, including \textsc{ChatTester}, \textsc{ChatUniTest}, and individual LLMs, by incorporating NLDs into their original frameworks. These adapted versions are denoted with an asterisk ($^*$), such as \textsc{ChatTester}$^*$ and \textsc{ChatUniTest}$^*$. By comparing \textsc{DISTINCT} with these NLD-integrated variants, we ensure evaluation fairness and specifically examine the effectiveness of our proposed architecture. We exclude \textsc{EvoSuite} and certain other LLM-based testing tools, such as \cite{TestGen,HITs,nan2025test}, for comparison, as their frameworks cannot accommodate NLDs, rendering them unsuitable for the non-regression testing scenario in our setting. Additionally, the work of Liu et al. \cite{AID} is excluded because their experimental setup requires both NLDs and a series of extra test cases for detecting specific tricky bugs, which differs significantly from our focus on generating tests solely based on natural language descriptions and defective code. Furthermore, we exclude standalone oracle generation methods (e.g., \cite{dinella2022toga,hossain2024togll}) for comparison, as they focus solely on assertion logic rather than constructing entire runnable test suites, so these partial generation techniques are outside our scope.

\textbf {RQ3 (Ablation Study): How much does each component contribute to \textsc{DISTINCT}?}
To evaluate the contribution of each component within \textsc{DISTINCT}, we design five experimental variants, each representing an independent ablation of the full framework:  (1) \textsc{DISTINCT}-Ana: a variant of \textsc{DISTINCT} without the Analyzer component, primarily enhancing its ability by repairing test cases that fail to compile. (2) \textsc{DISTINCT}-Code is a variant that removes the focal method from the Analyzer, meaning the logical branch analysis is only based on the NLD, not the actual code. (3) \textsc{DISTINCT}-Val is a variant that removes the Validator component. (4) Individual LLM*,  using a basic prompt as Figure \ref{fig:generateprompt}, incorporating the NLD but without involving the Validator or Analyzer components. (5) \textsc{DISTINCT} is the complete framework, integrating both Validator and Analyzer components.

\textbf{RQ4 (LLM Generality): How does \textsc{DISTINCT} perform with different LLMs?} 
To assess the generality of our proposed approach across different LLMs, we replace the original LLM, namely DeepSeek, with three representative alternatives, i.e., LLaMA3-70B, LLaMA3-8B, and CodeLLaMA-7B. All are state-of-the-art LLMs released recently, covering different sizes and both general and code LLMs.

\textbf{RQ5 (Sensitivity Analysis): How do different settings of hyperparameters affect the performance of \textsc{DISTINCT}?} 
To investigate how adjusting the hyperparameters within the \textsc{DISTINCT} framework influences its performance, we employ a coarse grid of $\{$3, 5, 7, 9$\}$ for the iteration numbers of the Validator and Analyzer, which is chosen based on empirical observations.  This setting spans low-to-high budgets with constant steps, covers the practical range where refinement tends to saturate, and keeps the search space manageable. In addition, we discuss the performance variation of \textsc{DISTINCT} with different numbers of test cases (\textit{N}), i.e., \textit{N}=$\{$5, 6, 7, 8, 9, 10$\}$.

\section{Evaluation Results}
\subsection{RQ2 (Overall Effectiveness)}

\begin{table*}[htbp]
\centering
\footnotesize
\renewcommand{\arraystretch}{1.3}
\setlength{\abovecaptionskip}{1pt}
\setlength{\tabcolsep}{2pt}
\caption{Comparison of test case generation approaches.}
\resizebox{\textwidth}{!}{
\begin{tabular}{lcccccccccc}
\toprule
\multirow{2}{*}[-0.8em]{\textbf{Approaches}} &  \multicolumn{5}{c}{\textbf{Defects4J-Desc}} 
& \multicolumn{5}{c}{\textbf{QuixBugs-Desc}} \\
\cmidrule(lr){2-6} \cmidrule(lr){7-11}
& CSR& PR & DDR & BC & SC
& CSR & PR & DDR & BC & SC \\
\midrule

Individual LLM* & 22.40\% {\scriptsize (0.00\%)} & 10.18\% {\scriptsize ($\uparrow$87.13\%)} & 1.36\% {\scriptsize ($\uparrow\infty$)} & 66.70\% {\scriptsize ($\uparrow$14.21\%)} & 66.70\% {\scriptsize ($\downarrow$3.19\%)} & 75.00\% {\scriptsize (0.00\%)} & 42.50\% {\scriptsize ($\uparrow$13.33\%)} & 40.00\% {\scriptsize ($\uparrow\infty$)} & 87.46\% {\scriptsize ($\uparrow$0.84\%)} & 78.51\% {\scriptsize ($\uparrow$2.51\%)} \\

\textsc{ChatTester}* & 24.59\% {\scriptsize ($\uparrow$7.15\%)} & 13.93\% {\scriptsize ($\uparrow$324.70\%)} & 1.36\% {\scriptsize ($\uparrow\infty$)} & 71.27\% {\scriptsize ($\uparrow$10.33\%)} & 79.13\% {\scriptsize ($\downarrow$0.78\%)} & 77.50\% {\scriptsize (0.00\%)} & 60.00\% {\scriptsize ($\uparrow$14.29\%)} & 42.50\% {\scriptsize ($\uparrow\infty$)} & 87.80\% {\scriptsize ($\uparrow$0.17\%)} & 78.30\% {\scriptsize ($\uparrow$0.71\%)} \\

\textsc{ChatUniTest}* & 24.23\% {\scriptsize ($\uparrow$20.01\%)} & 14.49\% {\scriptsize ($\uparrow$95.81\%)} & 1.36\% {\scriptsize ($\uparrow\infty$)} & 72.50\% {\scriptsize ($\uparrow$24.36\%)} & 78.20\% {\scriptsize ($\uparrow$15.34\%)} & 75.00\% {\scriptsize (0.00\%)} & 57.50\% {\scriptsize ($\uparrow$15.00\%)} & 42.50\% {\scriptsize ($\uparrow\infty$)} & 87.70\% {\scriptsize ($\uparrow$0.40\%)} & 78.85\% {\scriptsize ($\uparrow$1.87\%)} \\

\textbf{\textsc{DISTINCT} (Ours)} & \textbf{29.41\%} {\scriptsize ($\uparrow$19.60\%)} & \textbf{15.84\%} {\scriptsize ($\uparrow$9.32\%)} & \textbf{3.39\%} {\scriptsize ($\uparrow$149.26\%)} & \textbf{77.10\%} {\scriptsize ($\uparrow$6.34\%)} & \textbf{81.40\%} {\scriptsize ($\uparrow$2.87\%)} & \textbf{85.00\%} {\scriptsize ($\uparrow$9.68\%)} & \textbf{62.50\%} {\scriptsize ($\uparrow$4.17\%)} & \textbf{60.00\%} {\scriptsize ($\uparrow$41.18\%)} & \textbf{91.64\%} {\scriptsize ($\uparrow$4.37\%)} & \textbf{83.17\%} {\scriptsize ($\uparrow$5.48\%)} \\
\bottomrule
\end{tabular}
}
\vspace{-1mm}
\begin{minipage}{\textwidth}
\footnotesize
\vspace{2pt}
\raggedright
\textbf{Note:} Parentheses in the first three rows indicate improvement rates over their respective original versions (without NLDs), while parentheses in the last row indicate gains over the best-performing baseline.
\end{minipage}
\label{table2}
\end{table*}

Table \ref{table2} demonstrates the performance comparison among \textsc{DISTINCT} and the representative SOTA approaches across Defects4J-Desc and QuixBugs-Desc in the non-regression testing scenario. Here, the asterisk ($^*$) denotes the NLD-integrated variants of the respective baselines (i.e., \textsc{ChatTester}$^*$ and \textsc{ChatUniTest}$^*$). As can be seen, \textsc{DISTINCT} achieves notable improvements across all metrics, where it improves CSRs by 9.68\%--19.60\%, PR by 4.17\%--9.32\%, and DDR by 41.18\%--149.26\% across both datasets on average compared to the best-performing baselines. It also enhances BC by 4.37\%--6.34\%, and SC by 2.87\%--5.48\% on both datasets, demonstrating its superior effectiveness and robustness across benchmarks. Considering that both \textsc{DISTINCT} and SOTA approaches equipped with the validation-then-repair component (i.e., the Validator) and their only difference is the Analyzer component of \textsc{DISTINCT}, we attribute these improvements to the design of Analyzer, which leverages branch-consistency analysis guided by NLDs to explicitly direct the test case generation process, thereby significantly boosting the overall performance, especially defect detection rates. In addition, compared with the Individual LLM without either Validator or Analyzer, SOTA approaches and \textsc{DISTINCT} all manifest consistent improvements, which is reasonable.

Afterwards, we further make comparisons on existing approaches between their before- and after-adapted versions based on Table \ref{table1} and \ref{table2}. 
As can be seen, regardless of the SOTA approaches (e.g., \textsc{ChatTester}) or individual LLMs, they overall outperform their original versions, except for the slight performance decline of SC on individual LLM and \textsc{ChatTester}. For example, integrating NLDs into existing approaches brings average improvements of 0.00\%--20.01\% in terms of CSR, 13.33\%--324.70\% in terms of PR, and 0.17\%--24.36\% in terms of BC across both benchmarks. In particular, all approaches demonstrate preliminary defect detection capabilities in non-regression testing with a DDR score of 1.36\% on Defects4J-Desc and 40.00\%--42.50\% on QuixBugs-Desc.

\begin{tcolorbox}[colback=gray!10!white, colframe=gray!50!black, boxrule=0.5pt, before skip=5pt, after skip=5pt]
\textbf{\faPencil~Finding 2}: Incorporating NLD in the non-regression testing yields measurable gains to detect true bugs. \textsc{DISTINCT} substantially outperforms the LLM-based SOTA approaches across all evaluated metrics.
\end{tcolorbox}

\subsection{RQ3 (Ablation Study)}

RQ3 evaluates the contribution of each major component of \textsc{DISTINCT} to its overall performance. We conduct a systematic ablation study by assessing the four configurations mentioned in Section  \ref{Evaluation Setup}.
Table~\ref{table3} reports the performance of these variants and their deltas relative to the full system.

The Analyzer serves as the core component for aligning test generation with semantic intent. As shown in Table~\ref{table3}, removing this component (\textbf{\textsc{DISTINCT}-Ana}) leads to a severe degradation in defect detection, with DDR dropping by 39.82\% on Defects4J-Desc and 12.50\% on QuixBugs-Desc. Interestingly, on Defects4J-Desc, we observe a 4.08\% relative increase in CSR. This suggests that the Analyzer's NLD-guided consistency analysis often directs the LLM to explore complex code paths that are crucial for revealing deep-seated bugs. Without it, the model tends to generate more simplistic, syntactically safe tests, which significantly sacrifices DDR and PR (PR drops by 8.59\% and 4.00\% respectively).

Furthermore, the necessity of anchoring high-level intent to concrete code structures is evidenced by the \textbf{\textsc{DISTINCT}-Code} variant. Excluding code context from the Analyzer results in a notable decline in DDR (-26.55\%) and PR (-32.58\%) on Defects4J-Desc. This indicates that for large-scale Java projects, NLDs alone are insufficient; the code context is essential to ground high-level intent into concrete execution paths. In contrast, on the simpler QuixBugs-Desc benchmark, the impact is milder, though BC and SC still decline by 3.38\% and 1.78\%, respectively.

The Validator is equally critical for ensuring that semantically correct test ideas are translated into executable code. Removing the Validator (\textbf{\textsc{DISTINCT}-Val}) causes a sharp decline in CSR, with a relative drop of 12.27\% on Defects4J-Desc and 11.76\% on QuixBugs-Desc. This reduction in executability directly impairs the overall utility of the test suites, leading to a 19.76\% relative decrease in DDR on Defects4J-Desc and a 16.67\% decrease on QuixBugs-Desc. These results confirm that iterative repair based on execution feedback is indispensable for bridging the gap between initial reasoning and valid implementations.

Finally, comparing the full system to the \textbf{individual LLM*} reveals the cumulative impact of our design. The base configuration suffers catastrophic losses across all metrics, particularly in DDR (up to -59.88\%) and PR (up to -35.73\%). The fact that \textsc{DISTINCT} significantly outperforms all ablated variants across all coverage and correctness metrics demonstrates the synergy between components. The Analyzer identifies the correct semantic targets, the code context ensures these targets are grounded in the implementation, and the Validator ensures the final output is executable.

\begin{tcolorbox}[colback=gray!10!white, colframe=gray!50!black, boxrule=0.5pt, before skip=5pt, after skip=5pt]
\textbf{\faPencil~Finding 3}: 
Each module in \textsc{DISTINCT} is essential: the Analyzer is the primary driver for DDR, while the Validator is the main contributor to CSR. The synergy between NLD-guided analysis and iterative repair allows \textsc{DISTINCT} to effectively transform high-level intent into high-coverage, defect-revealing test suites.
\end{tcolorbox}

\begin{table*}[htbp]
\centering
\footnotesize
\renewcommand{\arraystretch}{1.3}
\setlength{\abovecaptionskip}{1pt}
\setlength{\tabcolsep}{3pt} 
\caption{Ablation study result of \textsc{DISTINCT}}
\resizebox{\textwidth}{!}{
\begin{tabular}{lcccccccccc} 
\toprule
\multirow{2}{*}{\textbf{Models}} 
& \multicolumn{5}{c}{\textbf{Defects4J-Desc}} 
& \multicolumn{5}{c}{\textbf{QuixBugs-Desc}} \\
\cmidrule(lr){2-6} \cmidrule(lr){7-11}
& CSR& PR & DDR & BC & SC
& CSR & PR & DDR & BC & SC \\
\midrule

\textsc{DISTINCT}-Ana & \textbf{30.61\%} {\scriptsize ($\uparrow$4.08\%)} & 14.48\% {\scriptsize ($\downarrow$8.59\%)} & 2.04\% {\scriptsize ($\downarrow$39.82\%)} & 75.40\% {\scriptsize ($\downarrow$2.20\%)} & 76.10\% {\scriptsize ($\downarrow$6.51\%)} & 82.50\% {\scriptsize ($\downarrow$2.94\%)} & 60.00\% {\scriptsize ($\downarrow$4.00\%)} & 52.50\% {\scriptsize ($\downarrow$12.50\%)} & 89.45\% {\scriptsize ($\downarrow$2.39\%)} & 81.84\% {\scriptsize ($\downarrow$1.60\%)} \\

\textsc{DISTINCT}-Code & 28.57\% {\scriptsize ($\downarrow$2.86\%)} & 10.68\% {\scriptsize ($\downarrow$32.58\%)} & 2.49\% {\scriptsize ($\downarrow$26.55\%)} & 75.90\% {\scriptsize ($\downarrow$1.56\%)} & 78.20\% {\scriptsize ($\downarrow$3.93\%)} & \textbf{85.00\%} {\scriptsize (0.00\%)} & \textbf{62.50\%} {\scriptsize (0.00\%)} & 57.50\% {\scriptsize ($\downarrow$4.17\%)} & 88.54\% {\scriptsize ($\downarrow$3.38\%)} & 81.69\% {\scriptsize ($\downarrow$1.78\%)} \\

\textsc{DISTINCT}-Val & 25.80\% {\scriptsize ($\downarrow$12.27\%)} & 15.64\% {\scriptsize ($\downarrow$1.26\%)} & 2.72\% {\scriptsize ($\downarrow$19.76\%)} & 75.20\% {\scriptsize ($\downarrow$2.46\%)} & 77.00\% {\scriptsize ($\downarrow$5.41\%)} & 75.00\% {\scriptsize ($\downarrow$11.76\%)} & 52.50\% {\scriptsize ($\downarrow$16.00\%)} & 50.00\% {\scriptsize ($\downarrow$16.67\%)} & 89.92\% {\scriptsize ($\downarrow$1.88\%)} & 81.54\% {\scriptsize ($\downarrow$1.96\%)} \\

individual LLM* & 22.40\% {\scriptsize ($\downarrow$23.84\%)} & 10.18\% {\scriptsize ($\downarrow$35.73\%)} & 1.36\% {\scriptsize ($\downarrow$59.88\%)} & 66.70\% {\scriptsize ($\downarrow$13.49\%)} & 66.70\% {\scriptsize ($\downarrow$18.06\%)} & 75.00\% {\scriptsize ($\downarrow$11.76\%)} & 42.50\% {\scriptsize ($\downarrow$32.00\%)} & 40.00\% {\scriptsize ($\downarrow$33.33\%)} & 87.46\% {\scriptsize ($\downarrow$4.56\%)} & 78.51\% {\scriptsize ($\downarrow$5.60\%)} \\
\midrule
\textbf{\textsc{DISTINCT}} & 29.41\% & \textbf{15.84\%} & \textbf{3.39\%} & \textbf{77.10\%} & \textbf{81.40\%} & \textbf{85.00\%} & \textbf{62.50\%} & \textbf{60.00\%} & \textbf{91.64\%} & \textbf{83.17\%} \\
\bottomrule
\end{tabular}}
\vspace{-1mm}
\begin{minipage}{\textwidth}
\footnotesize
\vspace{2pt}
\raggedright
\textbf{Note:} Values in parentheses indicate the relative differences, either increase ($\uparrow$) or decrease ($\downarrow$), to the performance of the full DISTINCT.
\end{minipage}
\label{table3}
\end{table*}
\begin{table*}[htbp]
\centering
\footnotesize
\renewcommand{\arraystretch}{1.3}
\setlength{\tabcolsep}{2.5pt}
\caption{Comparison of various models' performance across Defects4J and QuixBugs benchmarks.}
\vspace{-5mm}
\resizebox{\textwidth}{!}{
\begin{tabular}{lcccccccccc} 
\toprule
\multirow{2}{*}{\textbf{Models}} 
& \multicolumn{5}{c}{\textbf{Defects4J-Desc}} 
& \multicolumn{5}{c}{\textbf{QuixBugs-Desc}} \\
\cmidrule(lr){2-6} \cmidrule(lr){7-11}
& CSR& PR & DDR & BC & SC
& CSR & PR & DDR & BC & SC \\
\midrule
DeepSeek & 22.40\% & 10.18\% & 2.04\% & 66.70\% & 66.70\% & 75.00\% & 42.50\% & 40.00\% & 87.46\% & 78.51\% \\
\textbf{\textsc{DISTINCT}} & \textbf{29.48\%} {\scriptsize ($\uparrow$31.61\%)} & \textbf{15.84\%} {\scriptsize ($\uparrow$55.60\%)} & \textbf{3.39\%} {\scriptsize ($\uparrow$66.18\%)} & \textbf{74.20\%} {\scriptsize ($\uparrow$11.24\%)} & \textbf{81.40\%} {\scriptsize ($\uparrow$22.04\%)} & \textbf{85.00\%} {\scriptsize ($\uparrow$13.33\%)} & \textbf{62.50\%} {\scriptsize ($\uparrow$47.06\%)} & \textbf{60.00\%} {\scriptsize ($\uparrow$50.00\%)} & \textbf{91.64\%} {\scriptsize ($\uparrow$4.78\%)} & \textbf{83.17\%} {\scriptsize ($\uparrow$5.94\%)} \\
\midrule
Llama3-70B & 10.66\% & 7.03\% & 0.00\% & 58.40\% & 62.30\% & 70.00\% & 37.50\% & 32.50\% & 85.23\% & 76.32\% \\
\textbf{\textsc{DISTINCT}} & 26.98\% {\scriptsize ($\uparrow$153.10\%)} & 12.47\% {\scriptsize ($\uparrow$77.38\%)} & 3.17\% {\scriptsize ($\uparrow\infty$)} & 71.80\% {\scriptsize ($\uparrow$22.95\%)} & 80.30\% {\scriptsize ($\uparrow$28.89\%)} & 80.00\% {\scriptsize ($\uparrow$14.29\%)} & 55.00\% {\scriptsize ($\uparrow$46.67\%)} & 50.00\% {\scriptsize ($\uparrow$53.85\%)} & 89.47\% {\scriptsize ($\uparrow$4.97\%)} & 81.02\% {\scriptsize ($\uparrow$6.16\%)} \\
\midrule
Llama3-8B & 6.12\% & 5.57\% & 0.00\% & 69.50\% & 73.30\% & 65.00\% & 35.00\% & 30.00\% & 82.15\% & 74.18\% \\
\textbf{\textsc{DISTINCT}} & 19.73\% {\scriptsize ($\uparrow$222.39\%)} & 10.41\% {\scriptsize ($\uparrow$86.89\%)} & 2.95\% {\scriptsize ($\uparrow\infty$)} & 71.50\% {\scriptsize ($\uparrow$2.88\%)} & 80.40\% {\scriptsize ($\uparrow$9.69\%)} & 75.00\% {\scriptsize ($\uparrow$15.38\%)} & 50.00\% {\scriptsize ($\uparrow$42.86\%)} & 47.50\% {\scriptsize ($\uparrow$58.33\%)} & 87.09\% {\scriptsize ($\uparrow$6.01\%)} & 79.36\% {\scriptsize ($\uparrow$6.98\%)} \\
\midrule
CodeLlama-7B & 10.66\% & 5.44\% & 0.45\% & 58.40\% & 62.30\% & 65.00\% & 37.50\% & 32.50\% & 84.06\% & 75.21\% \\
\textbf{\textsc{DISTINCT}} & 21.09\% {\scriptsize ($\uparrow$97.84\%)} & 10.43\% {\scriptsize ($\uparrow$91.73\%)} & 2.95\% {\scriptsize ($\uparrow$555.56\%)} & 72.20\% {\scriptsize ($\uparrow$23.63\%)} & 80.90\% {\scriptsize ($\uparrow$29.86\%)} & 77.50\% {\scriptsize ($\uparrow$19.23\%)} & 52.50\% {\scriptsize ($\uparrow$40.00\%)} & 50.00\% {\scriptsize ($\uparrow$53.85\%)} & 88.53\% {\scriptsize ($\uparrow$5.32\%)} & 80.14\% {\scriptsize ($\uparrow$6.55\%)} \\
\bottomrule
\end{tabular}
}
\vspace{-1mm}
\begin{minipage}{\textwidth}
\footnotesize
\vspace{2pt}
\raggedright
\textbf{Note:} Values in parentheses represent the improvement rate relative to the baseline model.
\end{minipage}
\label{table4}
\end{table*}

\subsection{RQ4 (LLM Generality)}


Table~\ref{table4} summarizes the experimental results of \textsc{DISTINCT} integrated with various LLMs. For consistency, both the base models and \textsc{DISTINCT} were tested in the same NLD-based non-regressive setting; thus, the base model configurations are identical to those of the individual LLMs$^*$. Using DeepSeek as the backbone model, \textsc{DISTINCT} achieves the best performance. On Defects4J-Desc, it improves CSR from 22.40\% to 29.48\%, PR from 10.18\% to 15.84\%, and DDR from 2.04\% to 3.39\%. Code coverage also increases, with BC rising by 11.24\% and SC by 22.04\%. On QuixBugs-Desc, \textsc{DISTINCT} raises CSR to 85.00\% from 75.00\% and PR to 62.50\% from 42.50\%, while DDR improves from 40.00\% to 60.00\%.
The generality of \textsc{DISTINCT} is further validated across LLaMA3-70B, LLaMA3-8B, and CodeLLaMA-7B. 
As can be seen, \textsc{DISTINCT} improves Llama3-8B by 222.39\% in terms of CSR, by 86.89\% in terms of PR. Llama3-8B's DDR rises from 0\% to 2.95\%, and its coverage also saw a significant improvement.
In addition, Llama3-70B and CodeLlama-7B  both show substantial gains on \textsc{DISTINCT} compared to basic prompts as well. Specifically, Llama3-70B achieved a 153.1\% increase in CSP, a 77.38\% improvement in PR, and its DDR rose from 0\% to 3.17\%. CodeLlama-7B followed with a 97.84\% increase in CSP, a 91.73\% improvement in PR, and an impressive 555.56\% boost in DDR. Both LLMs also saw significant improvements in coverage.

While individual LLMs exhibit varied performance, \textsc{DISTINCT} with Llama3-8B underperforms those with medium-scale LLMs, such as Llama3-70B, and code-specific LLMs, such as CodeLlama-7B. This indicates that larger, more capable LLMs provide a stronger foundation for complex reasoning tasks \cite{kaplan2020scaling,wei2022emergent}, and that specialized pre-training on code can offer significant advantages in program understanding and test generation \cite{chen2021evaluating,roziere2023code}. Moreover, even these more powerful models fall short of the state-of-the-art DeepSeek in overall capability, underscoring the continued importance of raw model scale and advanced training methodologies. Despite these underlying differences in model capability, \textsc{DISTINCT} achieves consistent performance improvements across all these diverse backbones.

This consistent capability is a direct result of \textsc{DISTINCT}’s model-agnostic design. By integrating structured, external components, the framework provides crucial guidance that mitigates the inherent limitations of weaker or smaller LLMs. Specifically, the Analyzer refines test cases by enforcing branch consistency, and the Validator filters and repairs them based on execution signals. This comprehensive design effectively decouples test quality from the raw generative capabilities of the underlying LLMs, allowing \textsc{DISTINCT} to maintain high levels of Defect Detection Rate (DDR) and code coverage irrespective of the backbone model employed.

\begin{tcolorbox}[colback=gray!10!white, colframe=gray!50!black, boxrule=0.5pt, before skip=5pt, after skip=5pt]
\textbf{\faPencil~Finding 4}: \textsc{DISTINCT} generalizes well across LLMs of different scales and types, consistently improving key metrics including CSR, PR, DDR, and test coverage. 
\end{tcolorbox}

\begin{figure*}[h]
    \centering
    \includegraphics[width=1.0\textwidth]{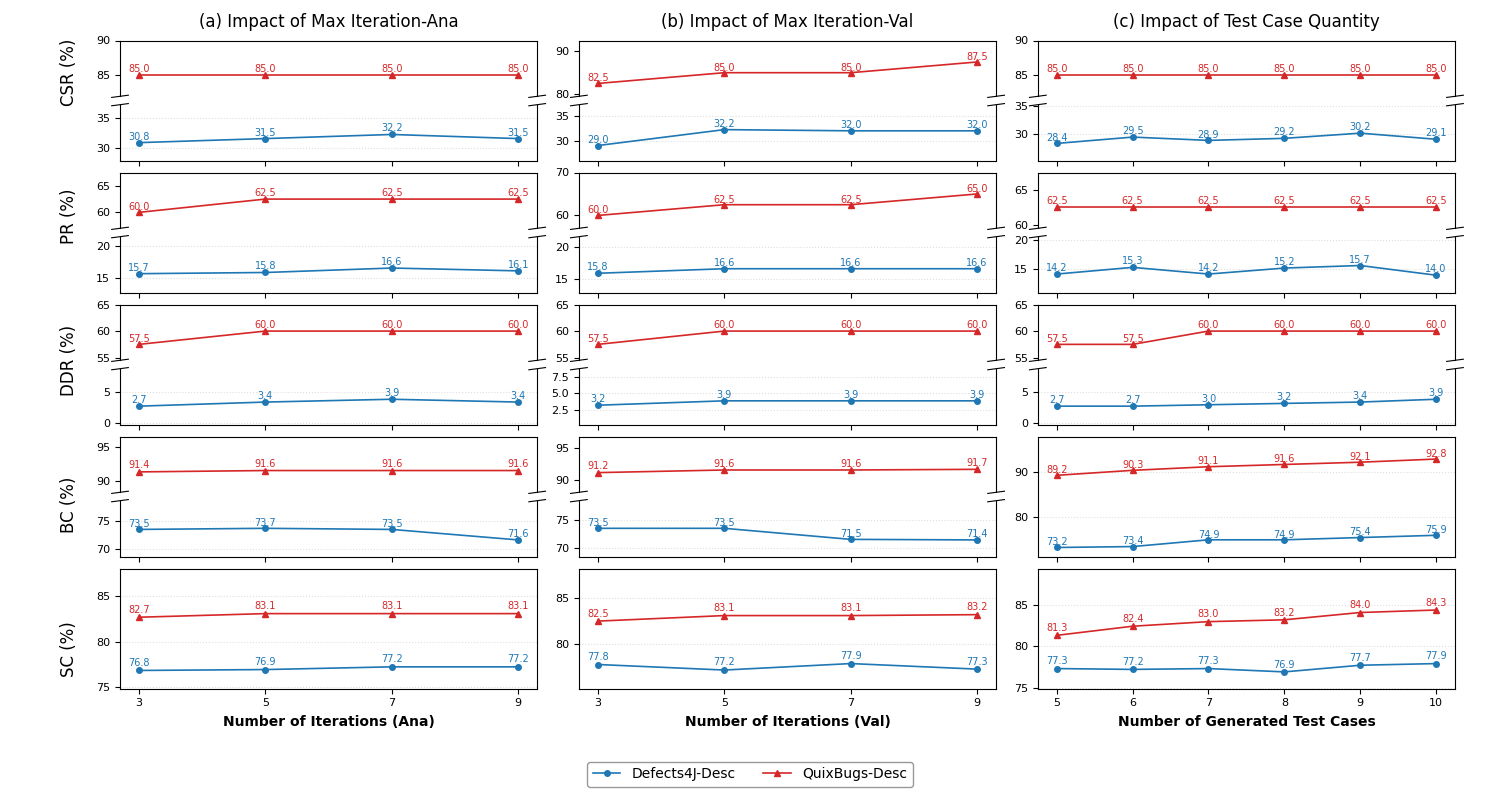} 
    \caption{Performance variation of \textsc{DISTINCT} with diverse hyperparameter settings}
    \label{fig:14}
\end{figure*}

\subsection{RQ5 (Sensitivity Analysis)}
\label{RQ5}

Figure~\ref{fig:14}(a)(b) shows the performance of our approach under varying iteration limits for the Validator (\textit{Max Iteration-Val}) and Analyzer (\textit{Max Iteration-Ana}) in non-regression testing on Defects4J-Desc and QuixBugs-Desc.
To cover the practical range for both hyperparameters and avoid insufficiency with low iterations and high cost with marginal gains on high iterations, we selected the following parameter ranges based on experience. 
To be specific, we first fixed the \textit{Max Iteration-Val} at 5, which is a reasonable initial value that has shown good performance in prior studies \cite{22}, and gradually increased \textit{Max Iteration-Ana} starting from the smaller iteration count of 3 to identify the best combination of \textit{Max Iteration-Val} and \textit{Max Iteration-Ana}. Subsequently, we fixed \textit{Max Iteration-Ana} at the value determined from the previous best combination, and then gradually increased \textit{Max Iteration-Val} from smaller iteration counts to investigate, across different datasets, which parameter settings enable the \textsc{DISTINCT} framework to achieve superior performance.

As shown in Figure~\ref{fig:14}(a), on Defects4J-Desc, when \textit{Max Iteration-Val} is fixed to 5, \textsc{DISTINCT} fluctuates slightly around 31\% in terms of CSR. 
Attributing to the iterative branch-consistency analysis within the Analyzer stage, 
\textsc{DISTINCT} not only prunes erroneous logical branches for generated test cases but also incorporates new ones inferred from the NLD to reveal defects progressively. This iterative process substantially improves the overall quality of the test suite, as reflected in the increase of PR, DDR, BC, and SC. In particular, DDR rises from 2.72\% to 3.85\%, indicating that repeated analysis is effective in strengthening defect-revealing capabilities.

However, when \textit{Max Iteration-Ana} increases from 7 to 9, performance decline on most metrics. A potential explanation is that most NLDs are not specific enough to explicitly cover all branches. 
As the number of iterations grows, generated tests increasingly conform to NLDs, leading to an over-concentration of test cases on NLD-related branches while reducing the diversity or even destroying other explored branches. Such an imbalance not only lowers branch coverage but also damages the test passing rates, ultimately decreasing the probability of exercising defect-triggering branches.

On QuixBugs-Desc, when \textit{Max Iteration-Val} is held constant to 5, increasing \textit{Max Iteration-Ana} does not affect CSR, which consistently remains at 85.00\%. Notably, when \textit{Max Iteration-Ana} increases from 3 to 5, the generated tests exhibit marked improvements across PR, DDR, BC, and SC. However, beyond five iterations, these performance indicators converge and remain stable. 
This can be explained by the characteristics of QuixBugs-Desc, where the defects are relatively prominent and can be effectively exposed with only a small number of iterations. Nevertheless, some methods under test involve invocations of external classes, which cannot be properly used within the generated tests. Consequently, for the vast majority of comparatively simple methods under test, a small and appropriate number of iterations is sufficient. However, there also exist a few methods for which the lack of complete contextual information causes the generated tests to either fail or miss defect detection, a limitation that is difficult to overcome even by increasing the number of iterations. These findings suggest that, for comparatively simple methods under test, the Analyzer is capable of achieving strong effectiveness with only a limited number of iterations.

Subsequently, we constrained \textit{Max Iteration-Ana} of Analyzer to 7 according to optimal values determined in the preceding experiments, thereby exploring the influence of \textit{Max Iteration-Val} in various values on \textsc{DISTINCT}.
As shown in Figure~\ref{fig:14}(b), the experimental results on Defects4J-Desc show that when \textit{Max Iteration-Val} increases from 3 to 7, almost all evaluation metrics improve.
Specifically, CSR improves by 10.17\%, PR rises by 4.48\%, and DDR boosts by 21.45\%, demonstrating the effectiveness of the repairing mechanism of Validator on the test case correction. However, when \textit{Max Iteration-Val} exceeds 7, 
\textsc{DISTINCT} stagnates on most evaluation metrics owing to the high difficulty of repairing certain cases.

On QuixBugs-Desc, the experimental results show that increasing the \textit{Max Iteration-Val} can, to some extent, improve the number of compilable tests, which in turn enables the Analyzer to generate more passing and defect-revealing tests. However, a larger \textit{Max Iteration-Val} does not necessarily lead to better performance of \textsc{DISTINCT}. Although increasing \textit{Max Iteration-Val} from 7 to 9 yields more compilable test suites, the limitations of the Analyzer prevent it from detecting the defects in these methods, so the DDR does not improve. These findings indicate that moderately raising the Validator’s iteration limit can enhance compilation success and thereby facilitate defect detection, but it also requires a higher Analyzer iteration limit, which comes at the cost of greater computational expense.

Overall, to achieve strong performance of \textsc{DISTINCT} across different datasets, it is necessary to appropriately tune \textit{Max Iteration-Val} and \textit{Max Iteration-Ana}. For complex datasets, \textsc{DISTINCT} usually performs well when both parameters are set to around 7, whereas for simpler datasets, fewer iterations are sufficient. Nevertheless, the optimal hyperparameter settings may still vary depending on the characteristics of each dataset.

After discussing the maximum iteration limits of both Analyzer and Validator, we turn to explore the influence of the number of generated test cases ($N$) on \textsc{DISTINCT}, where the specific evaluation results are shown in Figure~\ref{fig:14}(c).
For Defects4J-Desc, CSR and PR remain almost unchanged across different $N$, indicating that simply enlarging the test suite does not substantially affect compilability or executability. In contrast, DDR improves from 2.72\% at $N=5$ to 3.85\% at $N=10$. Branch coverage increases from 73.2\% to 75.9\%, while statement coverage rises from 77.3\% to 77.9\%.
These results suggest that on Defects4J-Desc, increasing the number of test cases mainly contributes to better coverage and defect detection rather than compilability. In practice, this implies that allocating a moderate number of test cases can effectively boost DDR and coverage, without needing to push CSR or PR further.
For QuixBugs-Desc, CSR and PR remain constant regardless of the value of $N$, where the former maintains a score of 85.0\% while the latter stays at 62.5\%. As $N$ continues to increase, both coverage metrics also show further improvements. This indicates that for QuixBugs-Desc, expanding the test suite beyond 7 cases per method still brings benefits in terms of coverage, even though the DDR remains stable. Practically, this suggests that increasing $N$ can enhance coverage, while the improvement in defect detection may plateau earlier.

Overall, these findings indicate that \textsc{DISTINCT} benefits from increasing the number of generated test cases, as this can improve both coverage and defect detection, although the degree of improvement may vary across metrics.

\begin{tcolorbox}[colback=gray!10!white, colframe=gray!50!black, boxrule=0.5pt, before skip=5pt, after skip=5pt]
\textbf{\faPencil~Finding 5}: 
Setting appropriate iteration limits for the Validator and Analyzer plays a critical role in balancing test quality and efficiency. Additionally, increasing the number of generated test cases tends to improve defect detection effectiveness, although the marginal gains in code coverage may diminish beyond a certain point.
\end{tcolorbox}

\section{Threats to Validity}
\textbf{Limited generality on Programming Languages (PLs).} This study only evaluated the test case performance of \textsc{DISTINCT} on Java programs without experiments on other mainstream PLs, such as Python or C++. However, Java is the most representative and widespread PL used in practice and most widely studied in the test case generation field \cite{HITs,tang2024chatgpt,22,23,wang2023jfinder}. As such, we believe the threat to this validity is limited. In the future, we will further assess the effectiveness of \textsc{DISTINCT} on other PLs. 

\textbf{Risk of Data Leakage.} Another potential threat to validity lies in the risk of data leakage from the benchmark datasets. Specifically, it is possible that data from Defects4J-Desc and QuixBugs-Desc may have been included in the training corpus of LLMs (e.g., DeepSeek), potentially leading to an overestimation of their capabilities in non-regression test generation. To mitigate this concern, we compared the test cases generated by \textsc{DISTINCT} with the official test suites of the datasets and found almost no overlap. This suggests that the LLM did not simply memorize existing tests. Moreover, a recent empirical study \cite{yang2025rethinking} demonstrated through extensive experiments that data leakage exerts minimal influence on code translation tasks, as their source and target codes are unpaired during LLMs’ pre-training. The test generation task is similar to code translation, as its focal methods (inputs) and corresponding test cases (outputs) lie in different files in the wild, thus also unpaired when pre-training. Therefore, from this aspect, we infer that the risk of data leakage in our setting is also minimal.

\textbf{Neutralness of the NLDs.} A further consideration is whether the constructed NLDs unintentionally reveal the underlying bugs, which could artificially inflate performance. To address this, we designed rigorous NLD construction protocols in Section \ref{Benchmark Construction} to follow, and strictly circumvented any reference to the specific buggy behavior or its location. 
This ensures that the LLM generates tests based on the intended specification rather than being guided by hints about the defect, thereby upholding the reliability of our evaluation in assessing true test generation capability.

\section{Background and Related Work}

\subsection{Unit Testing}

Unit testing serves as the primary verification technique to ensure that the implementation of a software component aligns with its intended logic\cite{olan2003unit,XIONG2025100313}. In practice, unit testing is categorized into two distinct paradigms based on the reliability of the code under test: regression testing \cite{wong1997study} and non-regression testing \cite{barr2014oracle}. The former assumes the focal method is logically sound and aims to detect future bugs during code evolution through regressions. In contrast, the latter focuses on just-in-time fault detection, where the test must serve as an independent oracle capable of revealing latent defects in the initial implementation \cite{barr2014oracle,konstantinou2024llms}. 


\subsection{Automated Unit Test Generation}

Writing high-quality tests is a time-intensive process. According to a survey by Daka et al.\cite{daka2014survey}, developers spend approximately 15\% of their working hours on writing test cases. To address this, a variety of automated testing tools have been developed in recent years. These tools aim to reduce the manual effort required for creating test cases while effectively uncovering potential software defects \cite{Myers2012}.

Traditional approaches such as Search-Based Software Testing (SBST) \cite{mcminn2004search} and symbolic execution \cite{king1976symbolic} treat generation as a coverage optimization problem. Tools such as \textsc{EvoSuite} \cite{fraser2011evosuite} and \textsc{Pynguin} \cite{lukasczyk2022pynguin} apply evolutionary algorithms to maximize code traversal. However, traditional approaches fundamentally rely on the existing code implementation as the primary source for search and generation. This dependency limits their ability to capture functional intentions that are not explicitly manifested in the code structure, often leading to test suites that fail to reveal latent defects within the original logic \cite{shamshiri2015automatically}.

Neural-based test generation initially focused on translating focal methods into assertions through Neural Machine Translation (NMT) architectures. \textsc{ATLAS} \cite{watson2020learning} first utilized NMT for assertion prediction, while subsequent retrieval-based models \cite{yu2022automated} incorporated contextual code examples to improve translation accuracy. The integration of Pre-trained Language Models (PLMs) such as BART and T5 shifted the focus toward end-to-end generation, where frameworks including \textsc{AthenaTest} \cite{tufano2020unit} and \textsc{TeCo} \cite{nie2023learning} fine-tuned transformer architectures on large-scale datasets to capture the mapping between code and tests. Neural-based test generation has primarily evolved within a regression testing paradigm, where the models are trained to mirror the existing implementation. NMT and PLM-based approaches rely on large-scale datasets of correct code, causing them to inherit a bias that treats the focal method as an absolute truth. Therefore, it cannot be adapted for the non-regression test generation.

Current state-of-the-art methods leverage LLMs through instruction-tuning\cite{YAO2024100211}. Tools like \textsc{ChatTester} \cite{Chattester} and \textsc{ChatUniTest} \cite{chen2024chatunitest} employ multi-turn dialogues and iterative repair to boost correctness. Advanced approaches like \textsc{CoverUp} \cite{altmayer2025coverup} and \textsc{SymPrompt} \cite{ryan2024code} further integrate static analysis to guide models toward higher coverage targets. \textsc{CodaMosa} \cite{lemieux2023codamosa} enhances SBST by using LLMs to provide high-quality seeds when traditional search reaches a plateau. Existing LLM-based approaches predominantly focus on achieving high coverage based on the provided implementation. Their prompts typically instruct models to verify the code as-is, treating the current implementation as the absolute ground truth. This prioritizes characterizing actual behavior over intended logic, leaving a critical gap in non-regression scenarios.

\section{Conclusion}
In this work, we present the first empirical study on unit test generation for non-regression testing, a common yet underexplored scenario in real-world development. We evaluate both traditional (\textsc{EvoSuite}) and LLM-based (\textsc{ChatTester}, \textsc{ChatUniTest}) approaches, finding that none can generate defect-revealing tests effectively. To address this limitation, we first construct two new benchmarks, Defects4J-Desc and Quixbugs-Desc, by augmenting the original benchmarks with NLDs. Then we propose \textsc{DISTINCT}, a novel LLM-based framework leveraging NLDs to guide the branch-consistency analysis, thereby revealing defects for focal methods.
Experiments on Defects4J-Desc and QuixBugs-Desc show that \textsc{DISTINCT} significantly improves DDR and other evaluation metrics, with robust generalization across different LLMs. These findings highlight the importance of description-guided and fault-aware test generation for improving practical software testing.

\section{Acknowledgement}
This work is supported by the National Natural Science Foundation of China (Grant Nos. 62502283, 62422208, 62232010, U24B20149, and 62302437), the Natural Science Foundation of Shandong Province (Grant Nos. ZR2024QF093 and ZR2025LZH006), and the Young Talent of Lifting Engineering for Science and Technology in Shandong, China (Grant No. SDAST2025QTB031), Ministry of Industry and Information Technology of China (Grant No. TC240A9ED-70), Research Project of Quancheng Laboratory, China (Grant No. QCL20250106), and Project of the Major Innovation Project of Key Laboratory of Computing Power Network and Information Security, Ministry of Education (Grant No. 2024ZD012).

\bibliographystyle{ieeetr} 
\bibliography{ref} 

\end{document}